\documentclass[sigconf,nonacm,screen]{acmart}

\setcopyright{none}
\renewcommand\footnotetextcopyrightpermission[1]{}

\usepackage{flushend}
\usepackage{tabularx}
\usepackage{amsmath}
\usepackage{amsfonts}
\usepackage{array}
\usepackage{booktabs}
\usepackage{multirow}
\usepackage{enumitem}
\usepackage{graphicx}
\usepackage{adjustbox}
\usepackage{subcaption}
\usepackage{xcolor}
\usepackage[breakable]{tcolorbox}
\usepackage{wasysym}
\usepackage{listings}
\usepackage{xurl}
\usepackage{xspace}
\usepackage[normalem]{ulem}
\usepackage{hyperref}
\hypersetup{hidelinks}

\lstdefinelanguage{Solidity}{
	keywords=[1]{anonymous, assembly, assert, balance, break, call, callcode, case, catch, class, constant, continue, constructor, contract, debugger, default, delegatecall, delete, do, else, emit, event, experimental, export, external, false, finally, for, function, gas, if, implements, import, in, indexed, instanceof, interface, internal, is, length, library, log0, log1, log2, log3, log4, memory, modifier, new, payable, pragma, private, protected, public, pure, push, require, return, returns, revert, selfdestruct, send, solidity, storage, struct, suicide, super, switch, then, this, throw, transfer, true, try, typeof, using, value, view, while, with, addmod, ecrecover, keccak256, mulmod, ripemd160, sha256, sha3},
	keywordstyle=[1]\color{blue}\bfseries,
	keywords=[2]{address, bool, byte, bytes, bytes1, bytes2, bytes3, bytes4, bytes5, bytes6, bytes7, bytes8, bytes9, bytes10, bytes11, bytes12, bytes13, bytes14, bytes15, bytes16, bytes17, bytes18, bytes19, bytes20, bytes21, bytes22, bytes23, bytes24, bytes25, bytes26, bytes27, bytes28, bytes29, bytes30, bytes31, bytes32, enum, int, int8, int16, int24, int32, int40, int48, int56, int64, int72, int80, int88, int96, int104, int112, int120, int128, int136, int144, int152, int160, int168, int176, int184, int192, int200, int208, int216, int224, int232, int240, int248, int256, mapping, string, uint, uint8, uint16, uint24, uint32, uint40, uint48, uint56, uint64, uint72, uint80, uint88, uint96, int104, uint112, uint120, uint128, uint136, uint144, uint152, uint160, uint168, uint176, uint184, uint192, uint200, uint208, uint216, uint224, uint232, uint240, uint248, uint256, var, void, ether, finney, szabo, wei, days, hours, minutes, seconds, weeks, years},
	keywordstyle=[2]\color{teal}\bfseries,
	keywords=[3]{block, blockhash, coinbase, difficulty, gaslimit, number, timestamp, msg, data, gas, sender, sig, value, now, tx, gasprice, origin, hookDelta, swapDelta},
	keywordstyle=[3]\color{orange!80!black}\bfseries,
	keywords=[4]{beforeSwap, afterSwap, beforeAddLiquidity, afterAddLiquidity, beforeRemoveLiquidity, afterRemoveLiquidity, beforeMint, afterMint, beforeBurn, afterBurn, beforeDonate, afterDonate, beforeInitialize, afterInitialize, onlyPoolManager},
	keywordstyle=[4]\color{magenta!80!black}\bfseries,
	identifierstyle=\color{black},
	sensitive=true,
	comment=[l]{//},
	morecomment=[s]{/*}{*/},
	commentstyle=\color{gray}\ttfamily,
	stringstyle=\color{red}\ttfamily,
	morestring=[b]',
	morestring=[b]"
}

\lstdefinestyle{solidity}{
	language=Solidity,
	extendedchars=true,
	basicstyle=\footnotesize\ttfamily,
	showstringspaces=false,
	showspaces=false,
	numbers=left,
	numberstyle=\footnotesize,
	numbersep=5pt,
	xleftmargin=1.5em,
	tabsize=2,
	breaklines=true,
	showtabs=false,
	captionpos=b,
	aboveskip=5pt,
	belowskip=5pt
}

\definecolor{rqboxbg}{RGB}{246,248,250}
\definecolor{rqboxline}{RGB}{83,110,135}

\newcommand{\rqanswer}[1]{\par
  \begingroup
  \setlength{\fboxsep}{4pt}\setlength{\fboxrule}{0.4pt}\noindent\fcolorbox{rqboxline}{rqboxbg}{\begin{minipage}{\dimexpr\columnwidth-2\fboxsep-2\fboxrule\relax}
    \normalsize\selectfont #1
    \end{minipage}}\par
  \endgroup
}

\begin{document}
\raggedbottom

\newcommand{\one}{({\em i}\/)\xspace}
\newcommand{\two}{({\em ii}\/)\xspace}
\newcommand{\three}{({\em iii}\/)\xspace}
\newcommand{\four}{({\em iv}\/)\xspace}
\newcommand{\five}{({\em v}\/)\xspace}
\newcommand{\six}{({\em vi}\/)\xspace}
\newcommand{\seven}{({\em vii}\/)\xspace}
\newcommand{\eight}{({\em viii}\/)\xspace}
\newcommand{\nine}{({\em ix}\/)\xspace}

\def\eg{\emph{e.g.}\xspace}
\def\etc{\emph{etc. }\xspace}
\def\ie{\emph{i.e.}\xspace}
\def\etal{\emph{et al.}\xspace}
\def\vs{\emph{vs. }\xspace}
\def\cf{\emph{cf. }\xspace}
\newcommand{\pb}[1]{\noindent{\bf \em #1}\hspace*{.3em}}

\title{An Empirical Study of Undisclosed Charges in Hook-Based Decentralized Exchanges}

\author{Kai Ma}
\affiliation{\institution{Huazhong University of Science and Technology}
  \country{Wuhan, China}
}
\email{kaima@hust.edu.cn}

\author{Tianle Sun}
\affiliation{\institution{Huazhong University of Science and Technology}
  \country{Wuhan, China}
}
\email{stlhust@hust.edu.cn}

\author{Shun Zhou}
\affiliation{\institution{Huazhong University of Science and Technology}
  \country{Wuhan, China}
}
\email{shunzhouaa@edu.cn}

\author{Boyao Yang}
\affiliation{\institution{Harbin Institute of Technology, Shenzhen}
  \country{Shenzhen, China}
}
\email{26b951041@stu.hit.edu.cn}

\author{Ningyu He}
\affiliation{\institution{Hong Kong Polytechnic University}
  \country{Hong Kong SAR, China}
}
\email{ningyu.he@polyu.edu.hk}

\author{Haoyu Wang}
\affiliation{\institution{Huazhong University of Science and Technology}
  \country{Wuhan, China}
}
\email{haoyuwang@hust.edu.cn}


\begin{abstract}
Many decentralized exchanges use automated market makers (AMMs) to trade assets against liquidity pools. Modern AMMs allow third-party contracts, called Hooks, to customize this process. However, Hooks can impose additional deductions that fee breakdowns based on standard pool parameters do not itemize. We present, to the best of our knowledge, the first large-scale empirical study of the production Hook ecosystem, focusing on these \emph{undisclosed settlement charges}. Our study covers 7,873 Hook deployments, 50,011 pools, and 296M pool events on Uniswap v4 and PancakeSwap Infinity across Ethereum and BNB Smart Chain (BSC) through August 1, 2026. We develop a bytecode analysis that reconstructs how Hooks change traders' payments and receipts, separating charges from exchanged assets and standard fees. We then reconstruct historical settlements to measure charging frequency and economic burden. We find that direct settlement adjustment is the dominant Hook capability, supported by 71.37\% of registered Hooks. Charging capability is widespread and often exercised: 24.30\% of Hooks can impose undisclosed deductions, and these deductions occur in 59.23\% of accounted swaps involving such Hooks. These charges generate millions of dollars in revenue, with \$3.97M recorded across Hooks with available token prices. These findings reveal a gap between standard fee displays and actual trading costs, motivating explicit disclosure of Hook deductions alongside standard fees.
\end{abstract}

\maketitle

\section{Introduction}

Decentralized exchanges (DEXs) enable users to trade crypto-assets through smart contracts. Many DEXs use automated market makers (AMMs), which execute trades against liquidity pools~\cite{asef2024x}. Hook-based AMMs allow external contracts called Hooks\footnote{Throughout this paper, the capitalized term ``Hook'' refers to the pool-customization contracts of Uniswap v4 and PancakeSwap Infinity.} to customize pool behavior. On August 1, 2026, DeFiLlama reports \$906.62M in total value locked (TVL) and \$526.96M in daily trading volume for Hook-based AMM protocols~\cite{hook-amm}, with Uniswap v4 and PancakeSwap Infinity accounting for over 98\% of this TVL~\cite{hook-amm,hook-amm-eth,hook-amm-bsc}. Hooks extend conventional AMM trading~\cite{asef2024x} with custom code executed before or after pool operations. This code can change pricing and settlement, including the amounts users pay or receive~\cite{uniswap-v4-hooks,pancakeswap-hooks}.

\begin{figure}[tbp]
    \centering
    \includegraphics[width=1\linewidth]{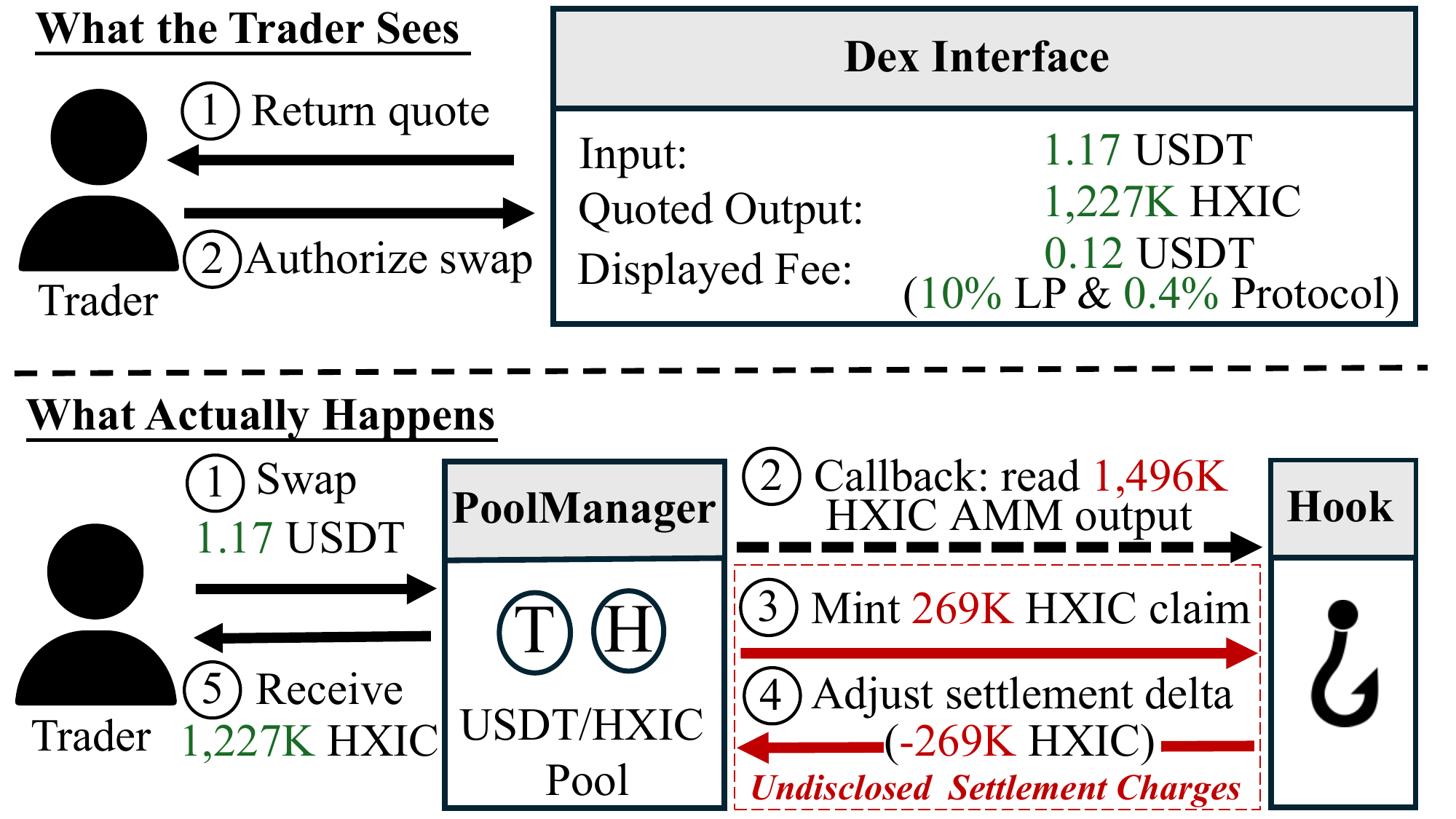}
    \caption{Traders receive the quoted output despite an unitemized charge. The negative value denotes reduced output.}
    \label{fig:undisclosed-charge-example}
\end{figure}

However, Hooks can use this control to impose additional charges beyond standard pool fees.
These deductions bypass the pool's standard fee parameters, so displays based on those parameters do not separately itemize them.
We refer to these unitemized deductions as \textit{\textbf{undisclosed settlement charges}}.
Figure~\ref{fig:undisclosed-charge-example} illustrates an undisclosed settlement charge through a real-world swap on the PancakeSwap interface~\cite{seaify-positive-slippage-hook,pancake-positive-slippage-hook-source}.
The interface quotes 1,227K HXIC for 1.17 USDT and displays only standard pool fees of 0.12 USDT (10\% liquidity-provider and 0.4\% protocol).
During settlement\footnote{BSC transaction \href{https://bscscan.com/tx/0xda0a75b79aa877337a3b3cffa9be9bfd883e83dfb573985a605e3dc4fc0ea6d8}{\texttt{0xda0a75...0ea6d8}}.}, the pool-management contract (PoolManager) deducts these fees and computes an output of 1,496K HXIC.
The Hook then retains 269K HXIC (18\%) of this output.
The trader therefore receives the quoted 1,227K HXIC, even though the fee breakdown does not itemize the Hook's additional charge.

Despite this visibility gap, existing studies on Hook-based AMMs focus on architectural overviews or general security evaluations~\cite{asef2024x,bachu2025overview,oh2025lifecycle}, leaving open how widespread these charges are, how often they occur, and what they cost traders.
Deployment counts alone do not reveal how widely Hooks are used, so interpreting charging prevalence also requires measuring their associated pools and transaction activity.
Identifying these charges then requires understanding deployed Hook behavior, often without verified source code.
It also requires reconstructing what users pay and receive to distinguish additional deductions from normal asset exchanges and standard pool fees.
We therefore combine bytecode analysis to identify charging mechanisms, pool configuration checks to assess whether they are permitted, and historical settlement reconstruction to measure their actual use and economic impact.

\noindent\textbf{This Work.}
We present, to the best of our knowledge, the first large-scale empirical study of the production Hook ecosystem in decentralized exchanges, with a focus on undisclosed settlement charges.
Our study covers 7,873 registered Hook deployments and 50,011 Hook-enabled pools on Uniswap v4 and PancakeSwap Infinity across Ethereum and BNB Smart Chain (BSC), through August 1, 2026.
We first characterize how Hooks are deployed and used, then examine charging mechanisms and their activation in practice, and finally quantify the resulting economic burden.
We organize our study around the following three research questions.

\begin{enumerate}[label=\textbf{RQ\arabic*.}, leftmargin=1.1em, itemsep=1pt, topsep=2pt, parsep=0pt]
    \item \textit{What characterizes the production Hook ecosystem in Hook-based AMMs?} 
    We systematically examine how deployment scale relates to economic activity, whether Hook rules are inspectable and stable, and which trading behaviors Hooks customize.
    
    \item \textit{How are undisclosed Hook charges distributed, enabled, and triggered?} 
    We empirically analyze charging mechanisms, compare their deployment footprint with recorded liquidity, and measure their occurrence across Hooks and historical swaps.
    
    \item \textit{What is the real-world economic burden of undisclosed Hook charges?} 
    Beyond cumulative revenue, we examine how charges are collected and provide fine-grained measurements of per-swap dollar and relative costs, comparing charging frequency and cost across platforms.
\end{enumerate}

\noindent\textbf{Our Contributions.}
\begin{itemize}[leftmargin=*, itemsep=1pt, topsep=2pt, parsep=0pt]
    \item \textbf{Charge identification.} We develop a bytecode analysis that reconstructs how Hooks change traders' payments and receipts, separating charges from exchanged assets and standard fees.

    \item \textbf{Ecosystem characterization (RQ1).} We characterize Hook deployment and usage, revealing that pool counts poorly reflect observed transaction activity (Spearman $\rho=0.145$).

    \item \textbf{Charging prevalence (RQ2).} We measure charging capability and its deployment footprint, revealing that 24.30\% of registered Hooks can impose additional charges. Their pools constitute 34.55\% of Hook-enabled pools but hold 2.69\% of recorded TVL.

    \item \textbf{Economic burden (RQ3).} We reconstruct historical trades and record \$3.97M in Hook revenue. Among accounted Uniswap v4 swaps involving charging-capable Hooks, charges occur more often on BSC than Ethereum (75.76\% vs.\ 43.53\%), yet mean charges per priced swap are lower (below \$0.41 vs.\ \$8.17).

\end{itemize}

\noindent\textbf{Paper Organization.}
Section~\ref{sec:background} introduces AMM settlement, Hook callbacks, and undisclosed Hook charges. Section~\ref{sec:methodology} presents our methodology. Section~\ref{sec:dataset-construction} describes dataset construction. Sections~\ref{sec:rq1}, \ref{sec:rq2}, and \ref{sec:rq3} present findings for \textbf{RQ1}--\textbf{RQ3}. Section~\ref{sec:discussion} discusses implications and limitations, and Section~\ref{sec:related-work} reviews related work.

\section{Background}
\label{sec:background}

\noindent\textbf{Hook-Based AMM Architecture.}
Hook-based AMMs separate pool execution from customization. Uniswap v4~\cite{uniswap-v4-hooks} and PancakeSwap Infinity~\cite{pancakeswap-hooks} use PoolManager contracts to maintain pool state and execute pool operations. Each pool can attach one Hook, and each Hook can serve multiple pools. To customize these operations, the PoolManager can invoke \emph{Hook callback functions}, such as \texttt{beforeSwap} and \texttt{afterSwap}, on the attached Hook before or after an operation (Table~\ref{tab:hook-insertion-points}). Pool permissions determine which callbacks run, while \emph{return-delta permissions} control whether returned adjustments change settlement. A Hook may execute through a proxy that delegates to a separate implementation contract. Upgrades can change this implementation while retaining the Hook address.

\begin{table}[tbp]
\centering
\caption{Hook callbacks by operation. Bold callbacks can adjust settlement. Liquidity callbacks differ between PancakeSwap Infinity concentrated-liquidity (CL) and Bin pools.}
\label{tab:hook-insertion-points}
\footnotesize
\begin{tabularx}{\linewidth}{@{}l>{\raggedright\arraybackslash}X@{}}
\toprule
Action (*: Bin) & Hook callbacks \\ \midrule
Initialize & \texttt{beforeInitialize}, \texttt{afterInitialize} \\
Add liquidity & \texttt{beforeAddLiquidity}, \textbf{\texttt{afterAddLiquidity}} \\
Remove liquidity & \texttt{beforeRemoveLiquidity}, \textbf{\texttt{afterRemoveLiquidity}}\\
Swap & \textbf{\texttt{beforeSwap}}, \textbf{\texttt{afterSwap}} \\
Donate & \texttt{beforeDonate}, \texttt{afterDonate} \\
Bin mint* & \texttt{beforeMint}, \textbf{\texttt{afterMint}} \\
Bin burn* & \texttt{beforeBurn}, \textbf{\texttt{afterBurn}} \\ \bottomrule
\end{tabularx}
\end{table}

\noindent\textbf{Swap Fees and Hook Charges.}\label{sec:bg-hooks}
Standard swap fees are collected within the pool's core exchange calculation. The pool charges fees on the input and exchanges the remainder. These fees compensate liquidity providers (LPs) and, when enabled, the protocol. In Listing~\ref{lst:bg-callback-return}, \texttt{\_swap} at L4 records this exchange in \texttt{swapDelta}, a pair of signed token amounts: negative for the user's payment and positive for the receipt~\cite{uniswap-v4-poolmanager}. For pools with dynamic LP fees, \texttt{beforeSwap} can set the LP fee rate for that swap through \texttt{lpFeeOverride} (L2). The core \texttt{\_swap} calculation uses this rate at L4, so the resulting fee remains included in \texttt{swapDelta}~\cite{uniswap-v4-ihooks,uniswap-v4-hooks-lib}.

Hooks can further adjust these exchange amounts through callback returns. Before the core exchange, \texttt{beforeSwap} also returns \texttt{deltaBefore} (L2), which can change the amount passed to the core swap (L3). Afterward, \texttt{afterSwap} inspects the core result and returns \texttt{deltaAfter} (L5). L6 combines these adjustments for each token into \texttt{hookDelta}, with positive amounts owed to the Hook and negative amounts supplied by it. L7 then subtracts this net Hook adjustment from \texttt{swapDelta} to obtain the user's final payment and receipt. A positive output-side \texttt{hookDelta} therefore \textit{reduces what the user receives and allows the Hook to retain the difference}.

\begin{lstlisting}[style=solidity,
firstnumber=1,
basicstyle=\scriptsize\ttfamily,
xleftmargin=0pt,
caption={Simplified swap flow with Hook adjustments.},label={lst:bg-callback-return}]
function swap(...) { // PoolManager
    (deltaBefore, lpFeeOverride) = hooks.beforeSwap(...);
    amountToSwap = adjustAmount(..., deltaBefore);
    swapDelta = _swap(..., amountToSwap, lpFeeOverride);
    deltaAfter = hooks.afterSwap(..., swapDelta, ...);
    hookDelta = combineDeltas(deltaBefore, deltaAfter);
    return swapDelta - hookDelta; }
\end{lstlisting}

\noindent\textbf{Undisclosed Settlement Charges.}\label{sec:bg-charges}
A Hook can also use settlement adjustments to receive input tokens and supply output tokens to execute a swap. Our study focuses on additional deductions imposed by Hooks beyond these exchanged token amounts and standard pool fees. Because these deductions bypass the standard pool-fee parameters, fee breakdowns based on those parameters omit them. We call such unitemized deductions \emph{undisclosed settlement charges}.

Listing~\ref{lst:hxic-charge} shows how the HXIC Hook retains output above the quoted amount in the trade from Figure~\ref{fig:undisclosed-charge-example}~\cite{pancake-positive-slippage-hook-source,seaify-positive-slippage-hook}. At L2 and L3, it reads the quoted exchange rate from the caller-supplied \texttt{hookData} and uses the input amount to compute the quoted output, \texttt{maxOut}, which is 1,227K HXIC in this trade. At L4, the core output \texttt{amountOut} of 1,496K HXIC exceeds this quote, yielding \texttt{excess} of 269K HXIC. At L5, \texttt{vault.mint} records 269K HXIC as a balance that the Hook can later withdraw from PancakeSwap Infinity's settlement contract. To charge the same amount to the user, L6 returns it as \texttt{deltaAfter}. The PoolManager includes this adjustment in \texttt{hookDelta} and subtracts it from the core output (Listing~\ref{lst:bg-callback-return}, L7). The user therefore receives the quoted 1,227K HXIC, while the standard fee breakdown omits the 269K HXIC charge.

\begin{lstlisting}[
style=solidity,
firstnumber=1,
basicstyle=\scriptsize\ttfamily,
xleftmargin=0pt,
caption={The HXIC Hook retains output above the quote.},
label={lst:hxic-charge}]
function afterSwap(...) external onlyPoolManager {
    (maxRate, denom) = HookDataDecoder.decodeExchangeRate(hookData);
    int256 maxOut = amountIn * maxRate / denom;
    int256 excess = amountOut > maxOut ? amountOut - maxOut : int256(0);
    if (excess > 0) vault.mint(address(this), currencyOut, uint256(excess));
    return (this.afterSwap.selector, int128(excess)); }
\end{lstlisting}

\section{Methodology \& Implementation}
\label{sec:methodology}
Section~\ref{sec:method-design} explains how we identify charging-capable Hooks, and Section~\ref{sec:implementation} describes the implementation.

\subsection{Methodology}
\label{sec:method-design}

\begin{figure}[tbp]
    \centering
    \includegraphics[width=1\columnwidth]{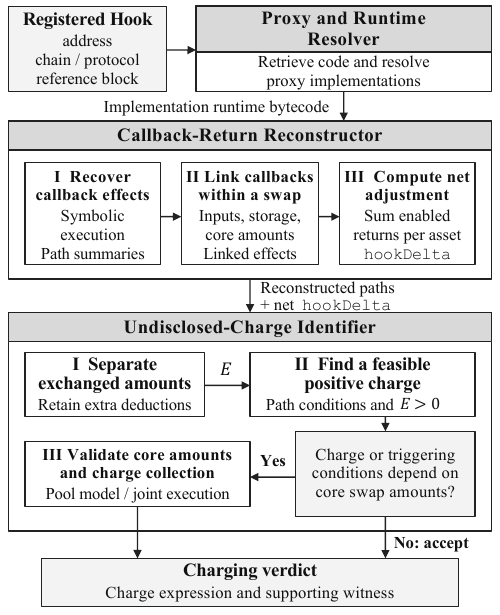}
    \caption{Charge identification ($E$: candidate charge).}
    \label{fig:charge-identification-workflow}
\end{figure}

\noindent\textbf{Method Overview.}
Our pipeline takes a registered Hook, its chain and protocol, and a reference block (Figure~\ref{fig:charge-identification-workflow}). First, the \emph{Proxy and Runtime Resolver} recovers the Hook's implementation bytecode. Second, the \emph{Callback-Return Reconstructor} combines callback effects under shared inputs and state into a net \texttt{hookDelta}. Third, the \emph{Undisclosed-Charge Identifier} separates charges from exchanged amounts, tests path feasibility, and, when needed, validates core swap amounts and charge collection. For each charging operation type (swap or liquidity), it returns a charge expression and a witness assigning inputs and state that satisfy the path constraints. This establishes charging capability under those constraints. We separately check deployed pool permissions and historical use (\S\ref{sec:rq2}).

\noindent\textbf{Proxy and Runtime Resolver.}
\label{sec:method-runtime}
For direct deployments, the resolver retrieves Hook runtime bytecode at the reference block. For proxies, it follows implementation addresses from bytecode, storage slots, or beacon contracts to recover the callback implementation (\S\ref{sec:implementation}). It passes this code to the Callback-Return Reconstructor.

\noindent\textbf{Callback-Return Reconstructor.}
\label{sec:method-return}
The reconstructor analyzes callbacks, Hook functions such as \texttt{beforeSwap} and \texttt{afterSwap} invoked by the PoolManager (\S\ref{sec:background}). Callbacks in one swap can share state and offset one another's adjustments, so their effects must be analyzed together. We first summarize individual callback paths, then link them through shared inputs, storage, and core swap amounts, and finally sum their adjustments per asset into net \texttt{hookDelta}.

\textit{Step I: Recover callback effects.}
The reconstructor locates the protocol-defined callbacks in Table~\ref{tab:hook-insertion-points} by their function selectors (call identifiers) and symbolically executes the resolved bytecode. For a proxy Hook, it executes the implementation contract's code while modeling reads and writes to the proxy's storage~\cite{ethereum-yellowpaper}. Starting from symbolic storage values, it records branch conditions, persistent and transaction-local (transient) storage writes, and external calls. At \texttt{RETURN}, it reconstructs the return buffer, checks it against the return format prescribed by the callback's application binary interface (ABI), and extracts the signed settlement adjustment. Each path summary retains the conditions, state updates, external calls, and returned adjustment. For example, the HXIC summary in Listing~\ref{lst:hxic-charge} links the condition $\texttt{amountOut} > \texttt{maxOut}$ (L4) to the \texttt{vault.mint} call for \texttt{excess} (L5) and the returned adjustment $\texttt{int128}(\texttt{amountOut} - \texttt{maxOut})$ (L6).

\textit{Step II: Link callbacks within a swap.}
For swaps, the reconstructor uses summaries of \texttt{beforeSwap}, \texttt{afterSwap}, or both, according to the protocol's invocation rules. Uninvoked callbacks contribute no path conditions or state writes. A single callback's summary is used directly, including for post-operation liquidity callbacks. When both swap callbacks run, the reconstructor pairs their summaries and assigns shared inputs, including swap parameters and \texttt{hookData}, the same symbolic variables despite their different ABI offsets. It substitutes values written by \texttt{beforeSwap} into \texttt{afterSwap}'s reads of the same persistent or transient storage slots, in both returned expressions and path conditions. Unwritten slots retain their initial symbolic values. For example, a fee $f$ stored by \texttt{beforeSwap} must be used wherever \texttt{afterSwap} reads that slot. The core swap introduces another dependency: \texttt{beforeSwap} can change the user-specified swap amount. The user fixes either the input amount (exact-input) or the output amount (exact-output), making that side the \emph{specified asset}. The reconstructor adds the \texttt{beforeSwap} adjustment for this asset to the original signed swap amount (Listing~\ref{lst:bg-callback-return}, L3). For the updated amount, direction, and swap mode, it assigns shared symbolic variables to the core input and output used in \texttt{afterSwap} expressions and conditions. Finally, it conjoins the linked conditions and discards incompatible pairs.

\textit{Step III: Compute the net settlement adjustment.}
The reconstructor sums only returned adjustments enabled by the protocol's return-delta permissions. An executed callback with this permission disabled contributes zero adjustment, but its path conditions and state writes remain in the linked summary. The standard-fee field \texttt{lpFeeOverride} is also excluded from the sum (Listing~\ref{lst:bg-callback-return}, \S\ref{sec:bg-hooks}). Using swap direction and mode, the reconstructor maps the specified asset and the other asset to the pool's two assets, \texttt{currency0} and \texttt{currency1}. It then sums adjustments per asset into the net \texttt{hookDelta} subtracted from the core \texttt{swapDelta}~\cite{uniswap-v4-poolmanager,uniswap-v4-ihooks}. It retains individual callback contributions alongside this net adjustment. For example, adjustments of $-f$ and $+f$ to the same asset cancel, even though the second callback returns a positive amount.

\noindent\textbf{Undisclosed-Charge Identifier.}
\label{sec:method-charge-identification}
The identifier searches swap and liquidity paths separately. Steps~I and~II separate exchanged amounts from candidate charges and test whether a candidate can be positive under the Hook's execution constraints. Step~III applies only when the charge or its triggering conditions depend on core swap amounts. It verifies that the pool can produce those amounts and the Hook collects the charge in the same execution. Each search stops at its first accepted path, which establishes charging capability for that operation type. It records required callbacks and return-delta permissions for comparison with pool configurations (\S\ref{sec:rq2}).

\textit{Step I: Separate exchanged amounts.}
The identifier separates token amounts exchanged through the Hook (\S\ref{sec:bg-charges}) from extra deductions, producing a candidate charge expression $E_{\pi,j}$ for asset $j$ on each reconstructed path $\pi$. For partial replacement, where the Hook executes only part of the exchange, it treats opposite-signed \texttt{beforeSwap} adjustments as exchanged amounts when the specified-asset adjustment depends on the user's swap amount. For example, in an exact-input swap of 1,000 units, adjustments of $+100$ input units and $-50$ output units mean that the Hook exchanges 100 input units for 50 output units, leaving 900 input units for the core. The positive 100 is therefore excluded as an exchanged amount. The identifier does not generally split fees embedded in such partial exchanges. Full replacement occurs when the \texttt{beforeSwap} adjustment cancels the user-specified amount passed to the core swap. For these paths, a positive adjustment to the other asset must match a supported fee form: an amount scaled by a nonnegative constant, or a residual $f$ after canceling a shared base amount $b$, as in $(b+f)-b$, with $f\geq 0$ established by path conditions. Positive \texttt{afterSwap} components are also retained separately: if \texttt{beforeSwap} supplies output $q$ via $-q$ and \texttt{afterSwap} deducts $f$, the candidate is $E_{\pi,j}=f$ even when the net adjustment $-q+f$ is negative. Other unmatched positive adjustments become candidates directly. These rules nominate candidates for the feasibility checks below.

\textit{Step II: Find a feasible positive charge.}
The identifier tests whether the candidate charge can be positive under the reconstructed Hook execution conditions. A positive $E_{\pi,j}$ denotes an extra deduction from the user's settlement. It therefore queries Z3 for
\begin{equation}
\operatorname{SAT}\left( \Phi_\pi \land \operatorname{signed}_{128}(E_{\pi,j}) > 0 \right),
\label{eq:positive-charge}
\end{equation}
where $\Phi_\pi$ combines linked callback conditions and state updates, non-revert conditions, return-value checks, and protocol-validity constraints. Pool fields, call input data (calldata), and initial storage remain symbolic within these constraints. The operator $\operatorname{signed}_{128}$ interprets the candidate amount as a signed 128-bit integer. For paths independent of core swap amounts, a satisfying assignment establishes charging capability. Otherwise, Step~III augments the same query before a verdict is accepted.

\textit{Step III: Validate core swap amounts and charge collection.}
Step~III checks that a positive candidate charge is consistent with the core exchange and its collection. Hook constraints alone may allow core amounts that the pool cannot produce. For Uniswap v4 exact-input paths with only \texttt{afterSwap}, the identifier computes core input and output amounts using a synthetic pool and the protocol's exact rounding rules~\cite{uniswap-v4-swapmath}. It substitutes these amounts into Equation~\eqref{eq:positive-charge} while keeping Hook storage symbolic. For paths whose cryptographic or cross-contract dependencies prevent a standalone positive witness, a historical swap involving the Hook provides code, pre-state, and signed calldata for joint execution of the PoolManager, vault, and Hook. Quote and signature data remain fixed while selected swap parameters vary symbolically. For HXIC, this binds \texttt{amountOut} in the charging condition and deduction to the PoolManager's computed output (Listing~\ref{lst:hxic-charge}, L4). Both execution models also require successful charge collection: \texttt{take} calls transfer tokens, or \texttt{mint} calls create redeemable claims for the Hook, with asset $j$ and amounts summing to $E_{\pi,j}$. Seeded execution additionally checks the corresponding increase in the Hook's retained vault balance. All constraints enter the same query, binding the charge, core amounts, and collection to one assignment.

\subsection{Implementation}
\label{sec:implementation}

We implement the pipeline in Python with Z3~\cite{demoura2008z3}, analyzing bytecode without requiring verified source code.

The \emph{Proxy and Runtime Resolver} extracts implementation addresses from bytecode for ERC-1167 minimal proxies~\cite{erc-1167} and clones with immutable arguments~\cite{openzeppelin-proxy-api}. For EIP-1967 Transparent, UUPS, and Beacon proxies~\cite{eip-1967,openzeppelin-proxy-api}, it reads standard implementation slots or queries beacon contracts.

The \emph{Callback-Return Reconstructor} uses a custom 256-bit symbolic Ethereum Virtual Machine (EVM) executor. Execution starts at a recovered callback entry. If no entry is recovered, it starts at the beginning of the bytecode with the callback selector fixed in calldata. Constants and ABI offsets remain concrete, while unknown calldata, initial persistent (\texttt{SLOAD}) and transient (\texttt{TLOAD})~\cite{eip-1153} storage values, environment variables, and external call returns are symbolic~\cite{ethereum-yellowpaper}. After state-changing calls, it discards cached storage reads to avoid reusing potentially outdated values. At \texttt{RETURN}, it checks the returned selector and extracts signed 128-bit settlement adjustments from the return buffer.

The \emph{Undisclosed-Charge Identifier} searches swap and liquidity paths with separate time limits. For paths accepted after Steps~I and II, \texttt{take} and \texttt{mint} calls provide additional collection evidence. Step~III uses seeded joint execution for core-dependent PancakeSwap Infinity CL paths on BSC when standalone analysis finds no positive witness. It accepts a solver assignment only if concrete re-execution reproduces the charge. Unmodeled, unvalidated, and timed-out paths remain unresolved.

\noindent\textbf{Identification Rules and Configuration.}
\label{app:charge-identification-config}
During expression normalization, an \texttt{int128} cast is removed only when path conditions prove that it preserves the value. Otherwise, the truncated expression remains unresolved. To recognize full replacement of the core exchange, let $a$ be the signed swap amount and $s$ the \texttt{beforeSwap} adjustment for the specified asset. The core receives $a_{\text{core}}=a+s$, so $s=-a$ cancels that exchange. Separately identified fees on the other asset and positive \texttt{afterSwap} deductions remain candidates.

\textit{Symbolic pool parameters.} Step~II restricts symbolic pool fields to protocol-valid values. Uniswap v4 requires currency ordering ($\texttt{currency0}<\texttt{currency1}$), tick spacing in $[1,32{,}767]$, and a static LP-fee field at most $10^6$ or the dynamic fee flag \texttt{0x800000}. PancakeSwap Infinity CL applies the same spacing and fee bounds, limits packed parameters to 40 bits, and constrains price limits by swap direction. A recovered pure permission getter's result constrains the callback registration bitmap. We also constrain Hook and caller addresses and callback and return-delta permissions.

\textit{Synthetic pool configuration.} For Uniswap v4 exact-input paths with only \texttt{afterSwap}, Step~III uses a pool at price one with input $10^6$ raw units, liquidity $10^{18}$, and zero protocol fee. It tests both swap directions with default LP fee zero and tick spacing 60, alongside bounded alternatives extracted from Hook guards. It computes core amounts under protocol rounding rules and substitutes them into the charge query, leaving Hook storage symbolic. Successful PoolManager \texttt{take} or \texttt{mint} calls must match the charged asset and a permitted recipient, with amounts summing to the charge.

\textit{Seeded joint execution.} For BSC Infinity CL paths requiring core amount validation, Step~III can jointly execute the PoolManager, vault, and Hook when standalone analysis finds no positive witness. An available historical seed supplies bytecode, pre-state, block context, and signed calldata. Quote and signature bytes remain fixed to preserve authorization, input-cap, and expiry guards. Selected swap amount words and the protocol fee field remain symbolic. The query ties the core and user settlement deltas to the charge, enforces the signed quote's guards, and requires a matching vault \texttt{mint} and an equal increase in the Hook's retained vault balance. Concrete re-execution must confirm the witness. For HXIC, valid witnesses retain the seeded swap amount with protocol-fee field values of 4,000 and 3,990. One extra raw unit fails authorization.
\section{Data Collection}
\label{sec:dataset-construction}

\begin{table}[tbp]
\centering
\caption{Dataset statistics to Aug.\ 1, 2026.}
\label{tab:dataset-snapshot}
\footnotesize
\setlength{\tabcolsep}{2pt}
\renewcommand{\arraystretch}{1.12}
\begin{adjustbox}{max width=\linewidth}
\begin{tabular}{@{}>{\raggedright\arraybackslash}p{0.27\linewidth}rrrrr@{}}
\toprule
& \textbf{Ethereum} & \multicolumn{3}{c}{\textbf{BSC}} & \\
\cmidrule(lr){2-2}\cmidrule(lr){3-5}
\textbf{Measure}
& \textbf{v4}
& \textbf{v4}
& \textbf{CL}
& \textbf{Bin}
& \shortstack{\textbf{Total}\\\textbf{unique}} \\
\midrule
Hook pools
& 27,231 & 15,310 & 7,388 & 82 & 50,011 \\
Registered Hooks
& 5,928 & 1,173 & 710 & 63 & 7,873 \\
\quad\textit{with verified source code}
& 2,467 & 186 & 309 & 40 & 3,001 \\
Pool activity events
& 5,463,709 & 1,860,250 & 288,707,011 & 83,227 & 296,114,197 \\
\quad\textit{Swap}
& 5,322,425 & 1,684,820 & 285,787,636 & 79,118 & 292,873,999 \\
\quad\textit{ModifyLiquidity}
& 128,974 & 161,090 & 2,858,062 & -- & 3,148,126 \\
\quad\textit{Donate}
& 12,310 & 14,340 & 61,313 & 0 & 87,963 \\
\quad\textit{Mint}
& -- & -- & -- & 2,905 & 2,905 \\
\quad\textit{Burn}
& -- & -- & -- & 1,204 & 1,204 \\
Transactions with receipts
& 4,213,904 & 27,138 & 2,659,236 & 8,154 & 6,907,077 \\
\quad\textit{unique senders}
& 288,983 & 1,401 & 17,857 & 1,691 & 309,186 \\
TVL in USD millions
& 42.848 & 4.474 & 66.928 & 0.011 & 114.260 \\
\bottomrule
\end{tabular}
\end{adjustbox}
\par
\parbox{\linewidth}{\scriptsize \textit{Note.} Totals count each address or transaction once per chain. BSC columns can share Hooks, transactions, or senders. The Mint and Burn rows count Bin liquidity operations.}
\par
\end{table}

Our dataset covers Uniswap v4 on Ethereum and BSC alongside PancakeSwap Infinity Concentrated Liquidity (CL) and Bin pools on BSC~\cite{hook-amm,hook-amm-bsc,hook-amm-eth}. We refer to these four configurations as platforms. Observation runs from each PoolManager deployment on January 23, 2025 for Ethereum v4, January 21, 2025 for BSC v4, and March 6, 2025 for Infinity through August 1, 2026. Pool initialization events identify 50,011 Hook pools. Their associated Hook addresses, deduplicated within each chain, yield 7,873 registered Hooks.

We collect activity events for these pools during the observation window and exclude initialization events from activity counts. Table~\ref{tab:dataset-snapshot} summarizes these events and the 6,907,077 successful transactions with receipts used to characterize the ecosystem (\S\ref{sec:rq1}). Call traces expose the nested contract calls needed to attribute Hook activity and reconstruct settlement. We retain these traces for 2,146,387 (31.08\%) of these transactions. Historical charge analysis in \S\ref{sec:rq2} and \S\ref{sec:rq3} also uses targeted trace collection for the 1,913 charging Hooks. The candidate set combines traced ecosystem transactions involving these Hooks with historical event and callback transactions retrieved from the deployment histories of the 1,913 charging Hooks, deduplicated by chain and transaction hash. All added candidates come from deployment histories rather than positive charge outcomes, using canonical decoded records through August 1, 2026 at Ethereum block 25,663,469 and BSC block 113,485,949. Across 2,175,181 candidate transactions, settlement reconstruction completes for 2,165,434 and leaves 9,747 without an analyzed result. Of the reconstructed transactions, 1,105,812 lie outside the receipt set in Table~\ref{tab:dataset-snapshot} because BSC receipt collection is incomplete. Because one transaction can contain multiple operations and trace reconstruction alone does not establish complete settlement accounting, we report operation accounting and pricing coverage in \S\ref{sec:rq2} and \S\ref{sec:rq3}.

At the cutoff, we retrieve runtime bytecode, pool configurations, and verified source code from block explorers. Pool TVL comes from Uniswap v4 subgraphs~\cite{thegraph-explorer} and the PancakeSwap Explorer API~\cite{pancakeswap-explorer-api}. We sum valid reported values for 49,378 of the 50,011 pools and treat missing or invalid values as unknown. Reported zeros may reflect unpriced tokens. The \$114.260M total covers Hook pools on Ethereum and BSC, while the \$906.62M protocol total on DeFiLlama includes pools without Hooks across all chains.

\section{RQ1: Hook Ecosystem Characterization}
\label{sec:rq1}

Across the four studied platforms~\cite{uniswap-v4-hooks,pancakeswap-hooks}, our dataset covers 7,873 registered Hooks governing 50,011 pools with \$114.26M in TVL (Table~\ref{tab:dataset-snapshot})~\cite{hook-amm,hook-amm-eth,hook-amm-bsc}. We examine where economic activity concentrates, whether Hook rules can be inspected and remain stable, and which trading behaviors Hooks customize.

\begin{figure}[tbp]
\centering
\includegraphics[width=0.9\linewidth]{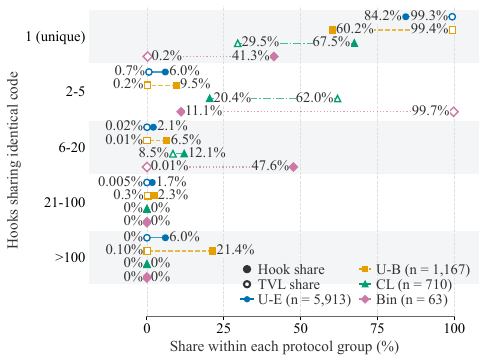}
\caption{Hook and TVL shares by the number of deployments sharing bytecode, normalized per platform. U-E and U-B denote Uniswap v4 on Ethereum and BSC.}
\label{fig:rq1-code-reuse}
\end{figure}

\noindent\textbf{Code Reuse, Liquidity, and Activity.}
The relationship between code reuse and liquidity differs across platforms (Figure~\ref{fig:rq1-code-reuse}). We recover implementation bytecode for 7,852 Hooks, identifying 6,463 distinct bytecodes, with 21.23\% of these Hooks sharing code with another deployment. In Uniswap v4, pools whose Hook bytecode appears in only one deployment hold 99.3\% of recorded TVL on Ethereum and 99.4\% on BSC. In contrast, pools whose Hook code appears in two to five deployments hold 62.0\% of TVL in PancakeSwap CL and 99.7\% in Bin.

\begin{figure}[tbp]
\centering
\includegraphics[width=\linewidth]{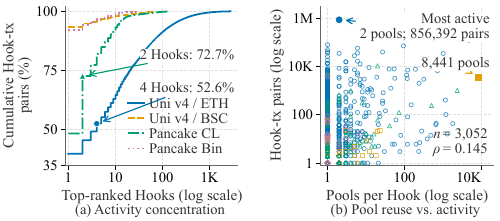}
\caption{Transactions across 3,052 observed Hooks, counted once per Hook: (a) cumulative shares with Hooks ranked by activity per platform and (b) counts versus attached pools.}
\label{fig:rq1-activity-concentration}
\end{figure}

Recorded transaction activity is concentrated in a few Hooks and only weakly related to pool counts. In the receipt-based traces (\S\ref{sec:dataset-construction}), excluding targeted charge backfill, 2,143,934 transactions have complete call trees with classified Hook calls, including failed internal calls, across 3,052 Hooks. Counting each transaction once per chain and invoked Hook yields 2,176,743 Hook transaction pairs, of which the top ten Hooks account for 63.81\%. Figure~\ref{fig:rq1-activity-concentration}a shows that activity is also concentrated within each platform. Transaction counts correlate weakly with attached pool counts (Spearman $\rho=0.145$, Figure~\ref{fig:rq1-activity-concentration}b). For example, Angstrom on Ethereum\footnote{\url{https://etherscan.io/address/0x0000000aa232009084bd71a5797d089aa4edfad4\#code}.} serves just two pools yet appears in 856,392 recorded transactions, whereas \texttt{ClankerHookStaticFeeV2} on BSC\footnote{\url{https://bscscan.com/address/0xc5d309026bcab6630888d51ce21154ad2f4828cc\#code}.} serves 8,441 pools but appears in only 3,569 transactions.

\textit{\textbf{Insight 1:} Pool counts are a poor guide to recorded transaction activity, with a correlation of only $\rho=0.145$ across the 3,052 Hooks in our call records.}

\noindent\textbf{Source Transparency and Upgradeability.}
Only 38.12\% of registered Hook addresses have verified source code on block explorers. Verification also differs across chains for widely reused code: among bytecodes deployed over 100 times, all 355 Ethereum Hook deployments are verified, but none of the 250 BSC deployments are.

Even when code is inspectable, a stable Hook address does not guarantee stable trading rules. Hooks can change these rules through upgrades, unlike conventional AMMs with immutable pool logic~\cite{asef2024x}. Of 533 proxy Hooks (6.77\%), 104 support replacing the implementation~\cite{eip-1967,openzeppelin-proxy-api}. On-chain histories record 49 implementation upgrades after pool creation across 14 Hooks serving 20 pools. Of these upgrades, 10 have source and runtime evidence of rule changes, including two with historical configuration evidence, 11 show controlled runtime differences, and 28 establish implementation replacement only. For example, the \texttt{MEVSwap} Hook\footnote{\url{https://etherscan.io/address/0x99d32f38aa4d1ec911420cba3b52d11cb9f0b0c0\#code}.} serves five active Ethereum pools. Its upgrades introduced a \texttt{tx.origin} blacklist that blocked 14 addresses and a check on ETH payments to the Hook after deducting base-fee gas costs. Raising the minimum from zero to 0.008 ETH then caused a payment of 0.001 ETH after this deduction to fail, although the prior version accepted it.

\begin{table*}[t]
\centering
\caption{Functional taxonomy of registered Hooks ($N=7{,}873$), including non-exclusive deployment shares and labeling criteria.}
\label{tab:rq1-functions}
\label{tab:functional-category-criteria}
\footnotesize
\setlength{\tabcolsep}{4.5pt}
\renewcommand{\arraystretch}{1.1}
\begin{tabularx}{\textwidth}{@{}>{\raggedright\arraybackslash}p{0.22\textwidth}rr>{\raggedright\arraybackslash}X>{\raggedright\arraybackslash}X@{}}
\toprule
\textbf{Domain / Sub-category} & \textbf{Hooks} & \textbf{Share} & \textbf{Inclusion criteria} & \textbf{Exclusions} \\
\midrule
\multicolumn{5}{@{}l}{\textbf{I. Settlement, Pricing \& Fee Customization}} \\[1pt]
\hspace{0.5em}Custom accounting & 5,619 & 71.37\% & Changes settlement through nonzero callback deltas or a protocol-linked claims ledger. & Ordinary token transfers, zero deltas, or asset withdrawals without a supported settlement path. \\
\hspace{0.5em}LP fee management & 484 & 6.15\% & Updates or overrides core LP fees through an applicable pool mechanism. & Fee calculations with no effective update, or overrides ignored by fixed-fee pools. \\
\hspace{0.5em}Rewards \& revenue distribution & 151 & 1.92\% & Accrues rewards or distributes rewards, rebates, or revenue to designated recipients. & Generic transfers, asset rescue, or game scores without reward logic. \\
\hspace{0.5em}Custom pricing & 104 & 1.32\% & Uses a custom quote or curve to determine amounts on a supported swap path. & Unused quotes, core-price reads, or LP fee changes alone. \\
\midrule
\multicolumn{5}{@{}l}{\textbf{II. Access, Launch \& Order Control}} \\[1pt]
\hspace{0.5em}Participation \& access control & 2,463 & 31.28\% & Restricts pool operations by participant identity, eligibility, timing, or launch state. & PoolManager authentication alone, generic argument checks, or fee exemptions. \\
\hspace{0.5em}Pool launch coordination & 234 & 2.97\% & Requests pool initialization or records a validated initialization lifecycle event. & Neutral initialization callbacks, or assumed liquidity provision by an external bootstrapper. \\
\hspace{0.5em}Order execution & 6 & 0.08\% & Maintains order state and processes fills or settlement on a supported execution path. & Order-related names or events without order-processing logic. \\
\midrule
\multicolumn{5}{@{}l}{\textbf{III. Liquidity \& External Capital Management}} \\[1pt]
\hspace{0.5em}External asset strategies & 428 & 5.44\% & Deposits, withdraws, lends, borrows, or stakes assets through an external strategy or vault. & An external address or call without a confirmed asset operation. \\
\hspace{0.5em}Liquidity management & 232 & 2.95\% & Creates, adjusts, or removes liquidity positions, including corresponding share accounting. & Pool initialization alone or a generic asset transfer. \\
\midrule
\multicolumn{5}{@{}l}{\textbf{IV. Telemetry \& External State Integration}} \\[1pt]
\hspace{0.5em}Application state integration & 126 & 1.60\% & Updates application state, content seeds, or an application engine from a swap path. & AMM accounting alone, administrative setters, or presumed payouts from game state. \\
\hspace{0.5em}Oracles \& market observation & 23 & 0.29\% & Maintains observations for a price or market-data service. & A price getter alone, or local swap statistics not serving an oracle or data feed. \\
\hspace{0.5em}Other identified functions & 59 & 0.75\% & Implements a verified function outside the named categories, such as local swap logging or statistics. & Unclear behavior, missing code, or an unverified external dependency. \\
\midrule
\textbf{Empty or uninitialized / No-op} & 770 / 21 & 9.78\% / 0.27\% & \multicolumn{2}{@{}l}{Empty bytecode, no callable callback logic under pool configuration, or selector-only neutral returns.} \\
\bottomrule
\end{tabularx}
\end{table*}

\noindent\textbf{Functional Landscape.}
\label{app:card-sorting}
\label{app:functional-classification}
Two authors with expertise in EVM bytecode analysis and DeFi developed our functional taxonomy through open card sorting~\cite{spencer2009card,zimmermann2016card}. Starting from 7,873 registered Hooks, they excluded deployments without substantive callback logic, derived functional categories from implementation evidence, and applied the resulting codebook to the remaining deployments.

\textit{Artifact Card Structure.}
We automatically generate an \emph{artifact card} for each of the 6,463 unique bytecodes recovered from 7,852 deployments. Each card summarizes callbacks, settlement and state effects, external calls, and guards, alongside source and trace evidence. Shared bytecode can behave differently across deployments, so the card also records their pool permissions, immutable arguments, and storage states. Verified source is available at 3,001 registered Hook addresses and for the resolved implementations of 2,958 deployments. Figure~\ref{fig:app-artifact-card} illustrates how the evidence supports a functional label for HXIC.

\begin{figure}[tbp]
\centering
\setlength{\fboxsep}{6pt}
\setlength{\fboxrule}{0.6pt}
\fbox{\begin{minipage}{0.94\linewidth}
\footnotesize
\textbf{HXIC artifact card} \\[-6pt]
\rule{\linewidth}{0.4pt}\\[2pt]
\textbf{1. Deployment:} BSC PancakeSwap CL. Direct implementation (HXIC Hook~\cite{pancake-positive-slippage-hook-source}). \\[3pt]
\textbf{2. Callbacks and permissions:} \texttt{beforeSwap} and \texttt{afterSwap} enabled. \texttt{afterSwapReturnsDelta = true}. \\[3pt]
\textbf{3. Conditions:} Input amount and signed quote rate in \texttt{hookData} set quoted output \texttt{maxOut}. Charging requires core output \texttt{amountOut} to exceed \texttt{maxOut}. \\[3pt]
\textbf{4. Settlement effects and external calls:} \texttt{beforeSwap} returns zero adjustment. \texttt{afterSwap} deducts the excess output, and \texttt{vault.mint} credits the same asset and amount to the Hook. No LP fee override, external strategy, or lending calls. \\[3pt]
\textbf{5. Evidence and label:} Verified source, recovered control flow, and sampled trace \texttt{0xda0a75...0ea6d8}. \\[1pt]
\emph{Settlement, Pricing \& Fee Customization} $\rightarrow$ \emph{Custom accounting}.
\end{minipage}}
\par
\caption{HXIC artifact card: an enabled output deduction and matching claim support the custom accounting label.}
\label{fig:app-artifact-card}
\end{figure}

\textit{Stage 1 (Baseline Filtering).}
We first exclude deployments without substantive callback logic at the August 1, 2026 cutoff. Of the 7,873 registered Hooks, 21 resolve to empty or uninitialized targets. Among the 7,852 code-resolved deployments, another 749 lack callable callback logic under their pool configurations, and 21 are no-op Hooks with selector-only returns. Excluding these 770 empty or uninitialized Hooks and 21 no-op Hooks leaves 7,082 deployments for functional classification.

\textit{Stage 2 (Codebook Development).}
The authors derive categories from the behaviors represented on the remaining cards. They inspect available source code and bytecode settlement or call summaries, assigning descriptive labels to callback behaviors without preset categories~\cite{zimmermann2016card}. Using AMM design concepts~\cite{uniswap-v4-whitepaper,uniswap-v4-vision,uniswap-v4-hooks,pancakeswap-vecake-exclusive-hook} and observed on-chain patterns, they merge synonymous labels into twelve sub-categories and group them into four domains. The codebook defines inclusion and exclusion criteria alongside deployment counts (Table~\ref{tab:rq1-functions}).

\textit{Stage 3 (Full-Dataset Labeling).}
Both authors independently apply the finalized codebook to all 7,082 remaining deployments. First, they label the capabilities implemented by each callback on a bytecode card. They then retain a label for an associated deployment only when its pool configuration enables the corresponding callback and any required return-delta permissions. This permits shared code review while distinguishing deployments with different permissions. All supported labels are retained, so a Hook can have multiple capabilities. A third author helps resolve disagreements by jointly reviewing control flow, available source, and historical traces until consensus is reached.

\textit{Labeling Criteria and Taxonomy Results.}
Each label in Table~\ref{tab:rq1-functions} requires implementation evidence for a reachable callback path under the deployment's configuration; function names or event signatures alone do not suffice. We apply these criteria to Hooks with and without verified source, so labels indicate supported capabilities rather than historical use. Registered Hooks fall into three groups: \emph{classified}, \emph{empty or uninitialized}, or \emph{no-op}. Classified Hooks may have multiple labels across enabled callbacks, and all percentages use the 7,873 registered Hooks. \emph{Other identified functions} requires verified callback logic outside the eleven named sub-categories.

The most common capability is directly changing the amounts users pay or receive at settlement. Although dynamic LP-fee adjustment is a flagship design motivation for Hook-enabled AMMs~\cite{uniswap-v4-hooks,pancakeswap-hooks}, only 6.15\% can adjust standard LP-fee rates (\emph{LP fee management}). By comparison, 71.37\% can directly adjust what users pay or receive at settlement (\emph{custom accounting}), and 31.28\% restrict participation or access. RQ2 examines when settlement adjustments become additional charges.

\textit{\textbf{Insight 2:} Directly adjusting users' settlement amounts is supported by 71.37\% of registered Hooks, compared with 6.15\% for adjusting standard LP fees.}

\rqanswer{\noindent\textbf{Answer to RQ1.}
Among registered Hooks, 71.37\% can directly change what users pay or receive at settlement, \vs 6.15\% that adjust standard LP fees. Pool counts weakly reflect transaction activity ($\rho=0.145$ across 3,052 Hooks in our call records).}
\section{RQ2: Undisclosed Charge Prevalence}
\label{sec:rq2}

We examine how settlement customization becomes a charging channel, how its deployment footprint relates to recorded liquidity, and how often charges occur in historical operations.

\noindent\textbf{Charging Prevalence and Mechanisms.}
\label{app:charge-identification}
Applying our static analyzer (\S\ref{sec:methodology}) to all 7,873 registered Hooks, we identify 1,913 deployments (24.30\%, spanning 1,427 unique bytecodes) capable of imposing undisclosed settlement charges. Two authors independently audited 100 static positives uniformly sampled from the 1,913 identified charging Hooks against the identification criteria (\S\ref{sec:method-charge-identification}), confirming all 100 as true positives (100\% sample precision).

\begin{table}[tbp]
\centering
\caption{Charge categories and matching historical evidence.}
\label{tab:rq2-charge-mechanisms}
\footnotesize
\setlength{\tabcolsep}{3.2pt}
\renewcommand{\arraystretch}{1.12}
\begin{adjustbox}{max width=\linewidth}
\begin{tabular}{@{}lrrrrr@{}}
\toprule
& \textbf{Static} & \multicolumn{4}{c}{\textbf{Historical Confirmation}} \\
\cmidrule(lr){2-2} \cmidrule(l){3-6}
\textbf{Charge mechanism} & \textbf{Hooks (\%)} & \textbf{Active ($H$)} & \textbf{Checked ($K$)} & \textbf{Charged ($F$)} & \textbf{Share ($F/K$)} \\ \midrule
Amount-based swap & 1,791 (93.62\%) & 1,487 & 1,457 & 871 & 59.78\% \\
Fixed-amount fee & 7 (0.37\%) & 5 & 4 & 2 & 50.00\% \\
Stored-amount fee & 1 (0.05\%) & 0 & 0 & 0 & -- \\
External-amount fee & 25 (1.31\%) & 22 & 15 & 1 & 6.67\% \\
Liquidity fee & 89 (4.65\%) & 57 & 57 & 2 & 3.51\% \\ \midrule
\textbf{Total} & \textbf{1,913 (100\%)} & \textbf{1,571} & \textbf{1,533} & \textbf{876} & \textbf{57.14\%} \\ \bottomrule
\end{tabular}
\end{adjustbox}
\par\smallskip
\begin{minipage}{\linewidth}
\scriptsize
Each Hook has one primary category. The liquidity row excludes 11 Hooks with detected charges in both operation types, which are assigned to swap categories. $H$: Hooks with category-relevant history. $K$: Hooks in $H$ with at least one operation whose settlement establishes a positive or zero charge. $F$: Hooks with observed charges matching their assigned category. $F/K$: share of checked Hooks with such charges.
\end{minipage}
\end{table}

Charging paths occur in both swap and liquidity operations. Searching each type separately (\S\ref{sec:method-charge-identification}) identifies swap charging in 1,824 Hooks and liquidity charging in 100, with 11 in both groups. To avoid double-counting, Table~\ref{tab:rq2-charge-mechanisms} assigns each Hook one primary category, prioritizing swap when both types are detected. For swaps, we classify the charge expression in the retained witness as amount-based when it depends on a trade amount, including through a stored or external rate. Otherwise, the category follows whether the charge amount is constant, stored, or externally supplied.

Charges tied to trade amounts dominate the assigned categories, accounting for 1,791 Hooks (93.62\%). Among these, 1,595 apply percentage cuts using constant (917) or stored (678) rates, 13 use external rates, and 183 have other expressions involving trade amounts, including partially recovered calculations.

\begin{table}[tbp]
\centering
\caption{Charging-capable Hooks: pool counts and TVL.}
\label{tab:rq2-prevalence}
\footnotesize
\setlength{\tabcolsep}{3.5pt}
\renewcommand{\arraystretch}{1.12}
\begin{adjustbox}{max width=\linewidth}
\begin{tabular}{@{}lrrr@{}}
\toprule
Protocol and chain
& \shortstack{All Hook-enabled\\pools' TVL}
& \shortstack{Attached\\pools' TVL}
& \shortstack{Attached pools with\\available TVL / total} \\ \midrule
Uniswap v4 ETH & \$42.848M & \$1.830M & 7,288/7,328 \\
Uniswap v4 BSC & \$4.474M & \$0.790M & 8,764/9,206 \\
PancakeSwap CL & \$66.928M & \$0.451M & 746/746 \\
PancakeSwap Bin & \$0.011M & \$0.000M & -- \\ \midrule
\textbf{Total} & \textbf{\$114.260M} & \textbf{\$3.072M}
& \textbf{16,798/17,280} \\ \bottomrule
\end{tabular}
\end{adjustbox}
\par
\end{table}

\noindent\textbf{Deployment Footprint and Liquidity.}
Charging capability spans a larger share of pools than of recorded liquidity. The 1,913 charging-capable Hooks serve 17,280 of 50,011 Hook-enabled pools (34.55\%). Table~\ref{tab:rq2-prevalence} reports \$3.072M in available TVL from 16,798 of these pools (2.69\% of recorded ecosystem TVL). Ethereum Uniswap v4 accounts for the largest dollar amount (\$1.830M), while BSC Uniswap v4 has the highest relative share (17.67\% of platform TVL).

\textit{\textbf{Insight 3:} Deployment counts alone cannot characterize charging exposure: pools attached to charging-capable Hooks constitute 34.55\% of Hook-enabled pools but hold only 2.69\% of recorded ecosystem TVL.}

\noindent\textbf{Charging in Practice.}
\label{app:historical-measurements}
Pool configurations permit the identified primary charging paths in 16,918 of the 17,280 associated pools (97.91\%, spanning 1,848 Hooks), enabling the required callbacks and settlement adjustments (\S\ref{sec:method-return}). The other 362 pools lack at least one required permission.

Historical reconstruction links each swap to its callback effects and resulting token movements to establish whether charging capability is exercised. After validating successful receipts and traces against the same transaction and block, we delimit the swap by its manager call path and identify its directly invoked callbacks using selectors, caller identities, and permissions. For swaps with a core AMM exchange, we match a unique core \texttt{Swap} event using the manager, pool, sender, amounts, and execution order, map returned adjustments to the two assets with disabled return-delta contributions discarded, and check that subtracting these adjustments from the core amounts reproduces the caller's settlement. For full-replacement swaps without a core exchange, settlement is reconciled directly from callback adjustments and token movements when exchanged amounts can be separated from extra deductions; ambiguous event associations or unexplained exchanged amounts remain unresolved.

We attribute transfers and protocol-claim changes to callbacks by asset, amount, and successful call path, checking receipt evidence and retained balances. Exchanged tokens, standard pool fees, refunds, unrelated payments, and later claim redemptions are excluded from new Hook revenue. An audited positive deduction establishes charging occurrence, while a zero-charge verdict requires the settlement to be fully explained by exchanged amounts and standard fees, including the effects of state-changing calls. Dollar-cost and ratio measurements in \S\ref{sec:rq3} additionally require every deduction to be fully matched to retained revenue, burns, or LP donations. The 361 swaps excluded from those measurements have valid core exchanges and settlement identities but incomplete matching of at least one deduction; they still count as charging swaps.

Of the 1,913 charging-capable Hooks, 1,571 have operations relevant to their primary category ($H$), and 1,533 have at least one with an established positive or zero charge ($K$, Table~\ref{tab:rq2-charge-mechanisms}). Matching executed Hook code and charge-producing expressions to the assigned category confirms charges in 876 of these 1,533 Hooks ($F$, 57.14\%).

A supplementary audit checks for historical swap deductions regardless of a Hook's primary category. For each combination of chain, Hook, callback selector, and return-value sign pattern, we select up to four observations ordered by SHA256 hash and add each Hook's latest swap. Across the 1,568 Hooks with swap history, this yields 1,953 operations in 1,604 transactions. Of these operations, 1,389 have confirmed deductions, 441 have no \texttt{hookDelta} charge, and 123 remain unresolved. The confirmed deductions span 1,118 transactions and 1,097 Hooks. These historical measurements differ in scope: Table~\ref{tab:rq2-charge-mechanisms} confirms category-matching charges in 876 Hooks, this sample confirms deductions regardless of category in 1,097, and \S\ref{sec:rq3} verifies collected revenue in 1,172 across all available traces. These three Hook sets share 794 deployments.

We then measure charging frequency across individual swaps, including deductions outside a Hook's primary category. Across 1,568 charging-capable Hooks with swap history, we identify 2,505,716 eligible swaps after excluding 20,997 service operations from 2,526,713 swap calls. The excluded operations mint UNI20 inscriptions, consuming the entire specified input as a service charge without an AMM exchange. A swap is \emph{accounted} when settlement evidence establishes whether it incurs an additional deduction beyond exchanged tokens and standard pool fees. This succeeds for 2,482,619 eligible swaps (99.08\%), with coverage of 98.27\% on Ethereum Uniswap v4, 99.45\% on BSC Uniswap v4, and 99.98\% on PancakeSwap CL. For the remaining 23,097, we cannot separate exchanged tokens from charges or reconstruct the effects of state-changing external calls.

Charges recur within these accounted swaps, but their frequency varies by platform. Overall, 1,470,344 of the 2,482,619 swaps (59.23\%) incur deductions. The rates are 75.76\% on BSC Uniswap v4, 69.79\% on PancakeSwap CL, and 43.53\% on Ethereum Uniswap v4. RQ3 examines the economic costs of these deductions.

\rqanswer{\noindent\textbf{Answer to RQ2.}
We identify charging capability in 1,913 of 7,873 Hooks (24.30\%). Their pools constitute 34.55\% of Hook-enabled pools but hold 2.69\% of recorded TVL. Among 2,482,619 accounted swaps involving identified charging-capable Hooks, 1,470,344 (59.23\%) incur deductions.}
\section{RQ3: Economic Burden Analysis}
\label{sec:rq3}

Static identification establishes charging capability, whereas historical accounting distinguishes the resulting user costs from the \emph{Hook revenue} retained by the contract. These quantities need not coincide because deductions extracted from settlement (\S\ref{sec:rq2}) may also fund burns or LP donations. Costs therefore comprise all deductions beyond exchanged amounts and standard pool fees, including burns and LP donations that do not become Hook revenue, while unrelated donations and standalone burns are excluded. Among 1,470,349 charging operations (1,470,344 swaps and five liquidity operations), 10,815 include token destruction and 26,689 include LP donations. These operations may also generate Hook revenue, but the burned or donated amounts are excluded from that revenue. Cost measurements require every deduction to be fully matched to revenue, burns, or LP donations, while revenue can be verified from attributable transfers or new protocol claims. We value tokens in USD using Chainlink price feeds at the execution block or market quotes from the prior 24h. For USD valuation, accepted quotes must be positive, match token decimals, and carry timestamps at or before execution within the permitted 24-hour window alongside provider checks.

\noindent\textbf{Hook Revenue and Concentration.}
Among 1,913 charging-capable Hooks, 1,172 collect charges in available traces. Revenue accounting records direct transfers and new protocol claims, counting each claim at creation and excluding later redemption from new revenue. The audited ledger, deduplicated by event identity, contains 1,489,847 accepted records: 810,764 transfers and 679,083 claims linked to their mint traces. Of the 1,172 Hooks, 1,044 have priced revenue totaling \$3.97M (\$3,965,281.97) across 1,244,343 transactions. Direct token transfers via \texttt{take} contribute 74.84\% (\$2.97M), and protocol claims created through \texttt{mint} contribute 25.16\% (\$1.00M). Hooks can later redeem these claims for tokens.

\textit{\textbf{Insight 4:} Revenue accounting must track protocol claims alongside token payments because 25.16\% of recorded Hook revenue accrues as newly minted claims.}

\begin{figure}[tbp]
\centering
\includegraphics[width=1\linewidth]{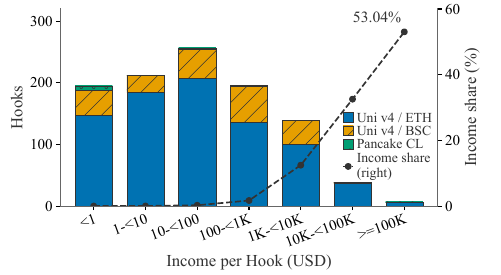}
\caption{Hook counts (bars) and revenue shares (line) by revenue band for 1,044 priced Hooks.}
\label{fig:rq3-income-detail}
\label{fig:rq3-income-bands}
\end{figure}

\begin{figure}[tbp]
\centering
\includegraphics[width=1\linewidth]{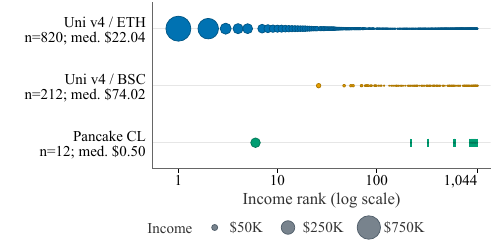}
\par
\caption{Global revenue ranks of 1,044 priced charging Hooks, grouped by platform. Bubble size indicates revenue.}
\label{fig:rq3-income-rank}
\end{figure}

\begin{figure}[tbp]
\centering
\includegraphics[width=1\linewidth]{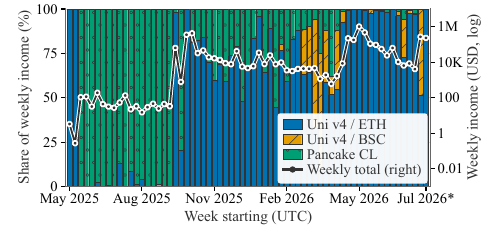}
\par
\caption{Weekly revenue: platform shares (bars, left) and USD totals (line, right, log scale). The final week (*) covers July 27 to August 1, 2026.}
\label{fig:rq3-income-weekly}
\end{figure}

Revenue is concentrated and sustained over time. Figure~\ref{fig:rq3-income-detail} shows that seven Hooks earning at least \$100K each account for 53.04\% of the total, and the top ten account for 59.02\%. Priced charges occur in 65 consecutive weeks from May 2025 through August 1, 2026. Figures~\ref{fig:rq3-income-rank} and~\ref{fig:rq3-income-weekly} show revenue across Hooks and over time: Hooks are ranked globally, while weekly bars report each platform's share alongside the total revenue curve.

\begin{figure}[tbp]
\centering
\captionsetup[subfigure]{font=footnotesize,justification=centering,singlelinecheck=false,skip=3pt}
\begin{subfigure}[b]{0.49\linewidth}
  \centering
  \includegraphics[width=\linewidth]{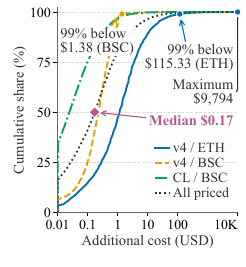}
  \caption{Additional cost (USD)}
  \label{fig:rq3-operation-usd}
\end{subfigure}\hfill
\begin{subfigure}[b]{0.49\linewidth}
  \centering
  \includegraphics[width=\linewidth]{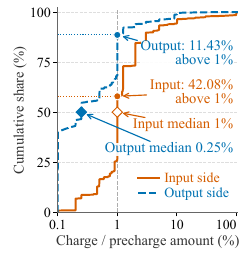}
  \caption{Charge ratio}
  \label{fig:rq3-charge-ratio}
\end{subfigure}
\caption{Cumulative distributions of Hook charges: (a) USD costs for 1,426,116 fully priced charging swaps and (b) fractions of the corresponding token amounts before deduction.}
\label{fig:rq3-operation-burden}
\end{figure}

\noindent\textbf{Per Swap Cost.}
Charging frequency and cost per swap differ across platforms. Of the 1,470,344 charging swaps in \S\ref{sec:rq2}, 1,426,116 (96.99\%) have fully reconciled deductions and accepted historical prices for every deducted asset (Figure~\ref{fig:rq3-operation-usd}). Coverage is 91.24\% on Ethereum Uniswap v4, 99.79\% on BSC Uniswap v4, and 99.95\% on PancakeSwap CL. The median additional cost is \$0.17, the mean \$2.79, and the maximum \$9,793.97 in one Ethereum swap. Mean costs are \$8.17 on Ethereum Uniswap v4 and below \$0.41 on BSC Uniswap v4 and PancakeSwap CL, despite more frequent charges.

\textit{\textbf{Insight 5:} Among identified charging-capable Hooks, charges occur more frequently on BSC Uniswap v4 than on Ethereum (75.76\% versus 43.53\% of accounted swaps), yet mean charges per fully priced charging swap are lower (below \$0.41 versus \$8.17).}

\noindent\textbf{Charges Relative to Swap Amounts.}
We measure charges relative to token amounts before Hook deductions (Figure~\ref{fig:rq3-charge-ratio}). The relative-cost analysis uses a narrower sample, requiring a valid core exchange taking one asset and supplying the other to define the precharge denominator, with the user paying at least that input and receiving at most that output. Differences must be fully matched to retained Hook revenue, burns, or LP donations. Hook-executed exchanges failing these checks are excluded. We divide each input deduction by core input plus that deduction, and each output deduction by core output before deduction. Core input includes standard pool fees. Both denominators exclude gas, and the input denominator excludes refunds. For example, in one Ethereum transaction\footnote{Ethereum transaction \href{https://etherscan.io/tx/0x7562ef289863adcec397f1445275d00638da669d6b15c5bde9ee667191fd0fc3}{\texttt{0x7562ef...fd0fc3}}.}, the user transfers $10^{13}$ wei (the smallest ETH denomination) but receives a refund leaving $5 \times 10^{10} + 2$ wei. The Hook deducts $5 \times 10^{10}$ wei, leaving $2$ wei for the core swap; the charge is nearly 100\% of the remaining input, but 0.5\% of the initial transfer.

Valid ratios cover 295,764 input-side and 1,177,962 output-side charging swaps, with 3,743 represented on both sides. The remaining 361 swaps have confirmed charges but lack complete reconciliation of every deducted amount, so they are excluded from both ratio and dollar-cost samples. The median ratios are 1.00\% on the input side and 0.25\% on the output side. Charges strictly exceed 1\% in 124,455 input observations (42.08\%) and 134,691 output observations (11.43\%). Deductions of this size can make narrow margin trades unprofitable without appearing in the quoted pool fee.

\noindent\textbf{Comparison with Standard Pool Fees.}
We compare Hook charges with standard LP and protocol fees in the same swaps at the same historical token prices. To recover exact standard fees, we replay the pool's swap calculation using \texttt{Pool} and \texttt{SwapMath} and reproduce the recorded core deltas, final price, liquidity, tick, and applied fees for 743 swaps: 735 on Ethereum Uniswap v4 and eight on PancakeSwap CL, seven of the latter with nonzero protocol fees. Of 456 replays with comparable prices, 154 have zero standard fees. In the remaining 302, Hook charges exceed standard fees in 299 (99.01\%).

To extend the comparison beyond exact replays without replaying every swap, we compute bounds on standard fees that account for the protocol's integer rounding. The core swap rounds fees up at each calculation step, so rounding increments can accumulate across a swap. Let $x_i$ be the gross core input consumed at step $i$, including its standard fee $t_i$, both in raw token units. Let $f$ be the total fee rate recorded in the \texttt{Swap} event, fixed during the core swap and expressed with denominator $D=10^6$. For $0<f<D$, protocol rounding gives $t_i=\lceil x_i f/D\rceil$. With total gross input $X=\sum_i x_i$, total standard fees $T=\sum_i t_i$, and at most $M$ steps, summing the rounded-up fees gives
\[
L=\left\lceil\frac{Xf}{D}\right\rceil,
\qquad
L\leq T\leq U=\min\{X,L+M-1\}.
\]
Each additional step can add at most one raw token unit to the rounding difference, and fees cannot exceed gross input $X$. We use the conservative step bound $M=2{,}000{,}000$, based on monotonic price progression through the protocol's finite tick range, including boundary and terminal steps. For $f=0$, both bounds are zero.

Each comparison is tied to a chain, transaction, Hook, manager call path, and block, using the input asset, $X$, $f$, token decimals, and accepted historical prices. We convert both bounds and Hook costs to USD under the same valuation policy. Costs strictly above the converted upper bound are classified as higher than the standard fee, and those strictly below the converted lower bound as lower. Across the 1,470,344 confirmed charging swaps, event records identify 728,986 with zero standard fees (49.58\%) and 741,358 with positive fee rates. Pricing Hook costs requires prices for the deducted assets, whereas comparing them with standard pool fees additionally requires a price for the core input asset; full pricing of Hook deductions therefore does not by itself establish price comparability for both fee types. Both fee types have comparable prices in 110,198 positive-rate swaps (14.86\%), while 631,160 lack them. The bounds establish higher Hook charges in 82,665 (75.01\%), lower charges in 27,497, and leave 36 indeterminate cases (which remain in the 110,198-swap denominator for the reported 75.01\%).

\textit{\textbf{Insight 6:} Hook charges can dominate standard pool fees: they exceed them in 82,665 of 110,198 charging swaps with positive pool-fee rates and comparable token prices (75.01\%).}

\rqanswer{\textbf{Answer to RQ3.}
Across 1,044 priced Hooks, recorded revenue totals \$3.97M, with 25.16\% accruing as protocol claims. Mean charges per fully priced charging swap are \$8.17 on Ethereum Uniswap v4 and below \$0.41 on BSC Uniswap v4. Hook charges exceed standard fees in 82,665 of 110,198 charging swaps with positive pool-fee rates and comparable prices (75.01\%).}
\section{Discussion}
\label{sec:discussion}

\noindent\textbf{Fee Disclosure and Monitoring.}
Fulfilling a quoted output does not establish fee transparency. In the HXIC example (Figure~\ref{fig:undisclosed-charge-example}), the trader receives the quoted tokens while the Hook retains 18\% of the core output, absent from the displayed fee breakdown. This gap arises because Hook deductions bypass standard pool-fee accounting, even when their rates are fixed. Dynamic charges further require disclosure to reflect the trade's inputs and execution state. Interfaces should therefore itemize standard fees and Hook deductions alongside expected payment and output. Pre-trade simulation could estimate these components for the trade's inputs and simulated state, with post-trade accounting reporting actual deductions. Both require separating charges from exchanged amounts carried in settlement adjustments (\S\ref{sec:method-charge-identification}). Monitoring should also track upgrades to detect trading-rule changes at existing Hook addresses.

\noindent\textbf{Interpreting Economic Exposure.}
Deployment counts alone cannot characterize exposure. Pool counts weakly track observed transactions (RQ1). Pools attached to charging-capable Hooks comprise 34.55\% of Hook-enabled pools but hold only 2.69\% of recorded TVL (RQ2). Higher charging frequency can also accompany lower per-swap costs (RQ3). Assessment and monitoring should therefore combine activity, charging frequency, and cost with deployment and liquidity measures. TVL measures capital at the cutoff, whereas cumulative charges span the observation window.

\noindent\textbf{Limitations and Scope of Inference.}
Measurements cover the four studied platforms through August 1, 2026. Charging rates use swaps with established charging status, while dollar totals use verified records with accepted historical prices and exclude unpriced charges. These subsets may differ systematically from omitted swaps, limiting generalization. Excluding unpriced charges does not guarantee that quotes are lower bounds, so \$3.97M denotes recorded revenue rather than a strict lower bound. Bounded execution and unresolved dependencies can miss charging paths, and one primary category cannot represent all mechanisms a Hook supports. The study measures deductions outside standard pool-fee accounting (\S\ref{sec:bg-charges}), without determining malicious intent or 
disclosure across all interfaces.

\section{Related Work}
\label{sec:related-work}

\noindent\textbf{Smart Contract Program Analysis.}
Static and symbolic analyzers inspect Solidity source code or recover execution paths and verify properties from EVM bytecode~\cite{luu2016making,tsankov2018securify,slither,mythril,grech2019gigahorse,sun2024all}, while fuzzers explore concrete state spaces~\cite{jiang2018contractfuzzer,shou2023ityfuzz}. More specialized analyses model fund flows to identify Ponzi schemes~\cite{chen2021sadponzi}, generate security-oriented descriptions for users~\cite{pan2023automated}, and detect allocation manipulation~\cite{pan2026allocscope}. Our analysis links shared inputs and state across Hook callbacks to reconstruct settlement effects and separate additional charges from exchanged amounts and standard fees.

\noindent\textbf{Smart Contract and DeFi Empirical Studies.}
Prior measurement studies characterize code reuse and diversity in the Ethereum contract ecosystem~\cite{he2020characterizing}. DeFi studies examine DEX execution, maximal extractable value (MEV)~\cite{daian2020flash,he2025gold}, flash-loan exploits~\cite{qin2021attacking}, price manipulation~\cite{kong2023defitainter,zhang2025following}, and traditional AMM mechanism design~\cite{werner2022sok,asef2024x}. Hook-enabled AMMs extend this setting by allowing external contracts to customize pricing and settlement. Our study characterizes how these contracts are deployed and reused, and measures the additional settlement costs they impose on traders.

\noindent\textbf{Hook-Enabled AMM Ecosystems.}
Documentation and early overviews describe Uniswap v4 Hook permissions and fee mechanisms~\cite{uniswap-v4-hooks,bachu2025overview}. Closest to our work, Oh and Lee~\cite{oh2025lifecycle} categorize structural threats and evaluate deployed Hooks through dynamic simulation. Our study measures the prevalence and economic burden of undisclosed settlement charges across Uniswap v4 and PancakeSwap Infinity, combining bytecode analysis with historical settlement reconstruction.

\section{Conclusion}
\label{sec:conclusion}

We combined bytecode analysis with historical settlement reconstruction to characterize the Hook ecosystem and measure undisclosed settlement charges on Uniswap v4 and PancakeSwap Infinity. Our analysis identified charging capability in 24.30\% of registered Hooks, while historical accounting recorded \$3.97M in Hook revenue. These charges bypass standard pool-fee accounting and can remain unitemized even when a trader receives the quoted output, showing that quote fulfillment alone does not establish fee transparency. Fee disclosure should therefore itemize additional Hook deductions alongside standard fees and reconcile the quoted breakdown with actual settlement.

\newpage
\bibliographystyle{ACM-Reference-Format}
\bibliography{ref}

\end{document}